\documentclass[aps,floats,preprint,superscriptaddress,floatfix]{revtex4}
\usepackage{epsfig}
\usepackage{amsmath}
\begin{document}
%\draft
\preprint{INFNCA-TH0206}
\title{Ultra high energy photon showers in magnetic field:
angular distribution of produced particles.
      }
\author{Massimo Coraddu}
\email{massimo.coraddu@ca.infn.it}
\affiliation{Dipart. di Fisica dell'Universit\`a di Cagliari,
             S.P. Sestu Km~1, I-09042 Monserrato (CA), Italy}
\affiliation{Ist. Naz. Fisica Nucleare (I.N.F.N.) Cagliari, 
             S.P. Sestu Km~1, I-09042 Monserrato (CA), Italy}
\author{Marcello Lissia}
\email{marcello.lissia@ca.infn.it}
\affiliation{Ist. Naz. Fisica Nucleare (I.N.F.N.) Cagliari, 
             S.P. Sestu Km~1, I-09042 Monserrato (CA), Italy}
\affiliation{Dipart. di Fisica dell'Universit\`a di Cagliari,
             S.P. Sestu Km~1, I-09042 Monserrato (CA), Italy}
\author{Giuseppe Mezzorani}
\email{giuseppe.mezzorani@ca.infn.it}
\affiliation{Dipart. di Fisica dell'Universit\`a di Cagliari,
             S.P. Sestu Km~1, I-09042 Monserrato (CA), Italy}
\affiliation{Ist. Naz. Fisica Nucleare (I.N.F.N.) Cagliari, 
             S.P. Sestu Km~1, I-09042 Monserrato (CA), Italy}
\date{October 7, 2002}
%\date{\today}
%
\begin{abstract}
Ultra high energy (UHE) photons can initiate electromagnetic showers in
magnetic field. We analyze the two processes that determine the
development of the shower, $e^+ e^-$ pair creation and synchrotron
radiation, and derive formulae for the angular distribution of
the produced particles. These formulae are necessary to study
the three-dimensional development of the shower.
\end{abstract}
%\pacs{}
\maketitle
%%%%%%%%%%%%%%%%%%%%%%%%%%%%%%%%%%%%%%%%%%%%%%%%%%%%%%%%%%%%%%%%%%%%%%
\section{Introduction}
\label{sect1}
Magnetic fields play a foundamental r\^ole not only for the acceleration and
propagation of charged cosmic rays, but also for the absorption of neutral
particles, photons and neutrinos, if the magnetic field is sufficiently strong
or the particles have sufficiently high energy~\cite{Ginzburg:sk}.
In particular, photons can initiate an electromagnetic shower in magnetic
field which is analogous to the showers produced in matter;
the main features of such showers 
(longitudinal development and particles energy spectra) 
has been analyzed in Refs.~\cite{Akhiezer:du,Anguelov:2000dz} 
under the assumption that momenta be orthogonal to the magnetic field.

Important examples of electromagnetic showers in strong magnetic fields
are the $\gamma$ and radio emission in pulsars~\cite{Sturrock:gp} and
blasars~\cite{Bednarek:1996py} in active galatic
nuclei~\cite{Bednarek:1998jq}. 
Strong magnetic fields of the order of $10^{12}$~G can be found in the
proximity of 
pulsars~\cite{Daugherty:mg,Da:82,Usov:1995tj,Baring:2000cr}:
in magnetic field of this order of magnitude photons loose energy by
radiating photons (photon splitting~\cite{Usov:2002df}) or creating
$e^+ e^-$ pairs~\cite{Daugherty:tr,Ba:88} which feed the cascade by producing
more bremsstrahlung photons. In such fields even neutrinos radiate
photons~\cite{Gvozdev:1996kx,Ioannisian:1996pn,Gvozdev:1997bs},
create $e^+ e^-$ pairs~\cite{Borisov:rb,Kuznetsov:1996vy,Gvozdev:1997bs}
or even $W$ particles~\cite{Erdas:2002wk}.

If the energy of the primary particle is sufficiently high, an
electromagnetic shower can develop even in weak fields, such as those present
in the interstellar medium or in the vicinity of stars and
planets~\cite{Ah:91,Vankov:ns}. A very 
important such a case is the shower produced by UHE ($E\gtrsim 10^{18}$~eV)
photons in the magnetic field outside the earth atmosphere: this early
shower influences the later atmospheric shower. 
These UHE photons are predicted by top-down theories 
as possible explaination of the experimental spectrum of ultra high 
energy cosmic 
rays~\cite{Berezinsky:1997td,Berezinsky:1997hy,Berezinsky:1997iz,%
Berezinsky:mw,Olinto:2002mc}).

In this paper we study the two processes that are the building block of
the electromagnetic shower in magnetic field: synchrotron radiation by
UHE electrons and $e^+ e^-$ pair production by 
UHE photons. In particular, we derive formulae for
the angular distribution. The precise dependence of the angular distribution 
from the field strength and from the energies of the particles is
necessary to determine important features of the phenomena, such as
the three-dimensional development and the lateral spreading.

A foundamental question is whether the three-dimensional development of
the shower could experimentally discriminate between UHE air showers
originated from a primary photon or from a primary proton
(or heavier hadron)~\cite{Ma:01,Vankov:ns,Bednarek:1999wg,Be:99a} 
and, therefore, discriminate between competing theories
of the high energy tail of the cosmic ray spectrum. In fact, UHE photons
start the shower well outside the atmosphere producing an additional
lateral spread to the subsequent atmospheric shower relative to a shower
originated by a proton. In addition a shower that begins outside the
atmosphere is less affected by the Landau-Pomeranchuk-Migdal
effect~\cite{Ka:96,Stanev:1996ux,Sh:01,Bednarek:2001fw}.

Another context where it is important the precise knowledge of
the angular distribution of the particles produced in the electromagnetic
shower is the modeling of the pulsar emission. For instance in the polar cup
model proposed by 
Sturrock~\cite{Sturrock:gp,Daugherty:mg,Usov:1995tj} high energy
electrons, due to the intense ($\sim 10^{12}\/ G$) magnetic fields, 
follow the field lines to minimize synchrotron radiation energy losses:
photons are emitted in a narrow cone nearly parallel to the field lines. 
The contribution of pair production to the photon interaction length depends
on the magnetic field component orthogonal to the photon momentum: since
momenta of the particles in the shower are nearly parallel to the field,
the precise emission angle might be critical for the shower development.

In the following Section~\ref{sect2} we introduce the notation, derive
formulae for the synchrotron radiation (magnetic bremsstrahlung) by
UHE electrons/positrons, discuss these formulae and show some
relevant plots. In Section~\ref{sect3} we make the analogous derivations
and discussion for $e^+ e^-$ pair creation by UHE photon in magnetic field.
The last Section~\ref{sect4} is reserved to our conclusions.

\section{Synchrotron radiation}
\label{sect2}
Synchrotron radiation (or magnetic bremsstrahlung) from ultrarelativistic
electrons has been studied by many authors, see for instance
Refs.~\cite{Ba:88,Erber:1966vv,Be:82,Schwinger:ix,Sokolov:nk}, 
where many results mediated over
the angular distribution can be found. For our study of the angular
dependence of both for the synchrotron radiation and the pair production
we shall follow the approach of 
Berestetskii-Lifshitz-Pitaevskii-Landau (BLPL)~\cite{Be:82}. In the following
discussion we shall assume that the electron momentum $\mathbf p$\/ is
perpendicular to the magnetic field $\mathbf H$\/: eventually we discuss
results for the general case in the last section (conclusions).

We recall some of the relevant notation. The characteristic parameter is  
\begin{equation} 
     \kappa =  \frac{H}{H_c} \frac{E}{mc^2} \quad ,
\label{eq:CharacPar}
\end{equation}
where $H$ is the stationary magnetic field, $m$ and $E$ the electron 
mass and energy and $ H_c$ is the critical field
\begin{equation} 
    H_c = \frac{m^2 c^3}{e \hbar} = 4.414\cdot 10^{13} \; \mbox{G } \quad , 
    \label{eq:CriticField}
\end{equation}
which is a natural quantum mechanical measure of the magnetic field
strength~\cite{Erber:1966vv}. The four-momenta of the incoming and outgoing
electron and the one of the emitted photon are, respectively,
$P=(E/c , \mathbf{p})$, $P'=(E'/c , \mathbf{p'})$\/ and
$K = (\hbar \omega /c , \mathbf{k})$. In most of the formulae we shall
use $\hbar = c = 1$.

We are interested in the limit when both the ingoing and outgoing electron
are ultrarelativistic, $E \gg m c^2$, and the field is relatively low
$H \ll H_c$. In this limit the quantization of the electron motion is not
important, since $E$ and $\Delta E=\hbar\omega$ are much larger than $\hbar
\omega_0 = \hbar c e H / E$, and we can use the semiclassical approximation
of BLPL with the electron following its classical orbit.

The differential probability to emit a photon is~\cite{Be:82}:
\[  
  dw = \sum_f \left| \int^{+\infty}_{-\infty}\, dt\,  V_{fi}(t) \right|^2
                 \frac{d^3 \mathbf k}{(2\pi)^3} 
\]
with
\[               
  V_{fi}(t) = e\int d^3 \mathbf r\; \psi_f^* \; 
               \vec{\alpha}\; \psi_i\; \vec{A}^* \, =\,
             \frac{e \sqrt{2\pi}}{\sqrt{\omega}} \langle f
            \left| Q(t) \right| i \rangle e^{i \omega t} \quad ,
\] 
were $\psi_{i,f}$ 
are the wave functions of the initial/final electron state,
$\vec{A}$\/ is the vector potential, and $ \vec{\alpha}$ are Dirac
matrices;  $Q(t)$ is the corresponding time-dependent Heisenberg operator.

Using the completeness relation  and
introducing the  variables $\tau = t_2 - t_1$  and $ t = (t_2 + t_1)/2$,
the probability of photon emission per unit time is:
\begin{equation}  
    \frac{dw}{dt} = \frac{e^2}{\omega} \frac{d^3  \mathbf k}{(2\pi)^2}
                 \int^{+\infty}_{-\infty}\, d\tau
    \langle i \left| Q^+(t+\tau /2)Q(t-\tau /2) \right| i \rangle 
     e^{-i \omega \tau} \quad ,
                 \label{eq:DiffProbSr} 
\end{equation}
where $d^3 \mathbf{k} = \omega^2 d \omega d \Omega = 
\omega^2 d \omega \sin\vartheta d\vartheta d\varphi$
with $\vartheta$\/ the angle between $\mathbf k$\/ and $\mathbf p$, 
and $\varphi$\/ the angle between $\mathbf H$\/ and the projection of 
$\mathbf k$\/  on the plane orthogonal to $\mathbf p$.

In Eq.~(\ref{eq:DiffProbSr}) the probability can be expanded in powers
of $\tau$, since the main contribution comes from small values of $\tau$,
when there is superposition between the amplitudes. In fact the values of
$\tau$ for which the superposition is significant can be evaluated using
semiclassical arguments. For kinematical reasons ultra-relativistic
electrons radiate in a narrow cone $\vartheta \lesssim m/E$: the amplitudes
for the emission sum coherently along a section of the classical electron
path where the direction changes of an angle $\vartheta \sim m/E$,
{\em i.e.}, $\omega_0 \tau \sim m/E$, giving a formation time
\begin{equation} 
   \tau_f \sim \frac{m}{E \omega_0} = \frac{m}{e H} =  
 \frac{1}{m}\frac{H_c}{H} =
  \frac{\hbar}{m c^2}\frac{H_c}{H}
          = 1.29 \cdot 10^{-21}  (H_c / H) \mathrm{sec}^{-1} \quad . 
                 \label{eq:FormationTime}
\end{equation}
To leading order in $\omega_0 \tau \sim m/E$ the resulting 
non-polarized photon emission probability is:
\begin{eqnarray}
 \frac{dw}{dt d\omega d\Omega} = && -\frac{e^2 \omega}{(2\pi)^2} 
      \int^{+\infty}_{-\infty}\, d\tau
      \left( \frac{E^2+{E'}^2}{4{E'}^2} (\omega_0 \tau)^2 + \frac{m^2}{E E'}
      \right)  \label{eq:EProbSrNp} \\
   && \times \exp{\left[ -i\frac{E}{E'}\omega\tau 
        \left( 1 - \mathbf n \cdot \mathbf v +\frac{\omega_0^2}{24}\tau^2  
       \right)
                 \right]}  \quad ,   \nonumber       
\end{eqnarray}
where $\mathbf{v}$ is the initial electron velocity.

The integration in $d\tau$ of Eq.~(\ref{eq:EProbSrNp}) yields
\begin{subequations}
\label{eq:AngDiffProbEmissSr}
\begin{equation}
 \frac{dw}{dt d\omega d\Omega} =
        \frac{e^2}{\pi} 
         \left[ \frac{E^2 + {E'}^2}{E E'} X 
                - \left( \frac{\omega}{\kappa E'} \right)^{2/3}
         \right] \Phi(X) \quad ,
\end{equation}
where
\begin{equation}
   X \equiv 2  \left(\frac{E}{m} \right)^2
        \left( 1\, -\,\left(1-\frac{m^2}{2E^2}\right)\cos\vartheta\right)
         \left( \frac{1}{\kappa} \frac{\omega}{E'} \right)^{\frac{2}{3}}
          \quad ,
\end{equation}
\end{subequations}
$\kappa$ is the characteristic parameter defined
in Eq.~(\ref{eq:CharacPar})
and $\Phi(X)$ is the Airy function~\cite{Ab:64}:
\begin{equation}
\Phi(x) = \frac{1}{\pi}
\int^{\infty}_{0} dt\,\cos\left( x t + \frac{t^3}{3} \right) \quad .
 \label{eq:DefAiryFunc}
\end{equation}

In the ultra-relativistic limit the differential
probability of photon emission,  Eqs.~(\ref{eq:AngDiffProbEmissSr}),
depends only on the angle $\vartheta$ between the electron and emitted
photon directions, threfore $d\Omega = 2 \pi d(\cos\vartheta)$.

It is often useful, for instance when writing the cascade equations, to
express the result in terms of the fractional energy carried by the photon 
$u = \omega /  E$;
then equations~(\ref{eq:AngDiffProbEmissSr}) become
\begin{subequations}
\label{eq:AngDiffProbEmissSru}
\begin{equation}
 \frac{dw}{dt\, d\hbar\omega\, d\cos\vartheta} =  
\frac{2 \alpha}{\hbar} 
      \left[\frac{1+ (1-u)^2}{(1-u)} X 
            - \left(\frac{u}{(1-u)\kappa}\right)^{2/3}
     \right] \Phi(X)
\end{equation}
with
\begin{eqnarray}
  X &=& 2  \left(\frac{E}{m} \right)^2
        \left( 1  - \left(1-\frac{m^2}{2E^2}\right)\cos\vartheta\right)
  \left(\frac{u}{(1-u)\kappa} \right)^{2/3}  \\
    &=& \left( 1+2(1-\cos\vartheta)\frac{E^2}{m^2} - (1-\cos\vartheta)\right)
      \left(\frac{u}{(1-u)\kappa} \right)^{2/3}
                 \label{eq:AngDiffProbEmissSrGenCaseFracEn}
\end{eqnarray}
\end{subequations}
where $\alpha$ is the fine-structure constant. In the ultrarelativistic
limit it is convenient to use $2(1-\cos\vartheta)\sim \vartheta^2 
\sim (m/E)^2$ as angular variable.

Equations~(\ref{eq:AngDiffProbEmissSr}) or, equivalently,
Eqs.~(\ref{eq:AngDiffProbEmissSru}) are our main new result, together 
with the analogous result for pair creation,
Eqs.~(\ref{eq:AngDiffProbEmissPpGenCaseFracEn}), result which is presented in
the next Section, {\em i.e.}, the angular dependence of the produced
particles.

The integral of Eq.~(\ref{eq:AngDiffProbEmissSru}) in $d\cos\vartheta$
yields the differential emission probability per unit energy:
\begin{subequations}
 \label{eq:DiffEnerProbEmissSr}
\begin{equation}
 \frac{dw}{dt d\omega} = - e^2 \left(\frac{m}{E}\right)^2 
                        \left( \int^{\infty}_{\xi} \Phi(x)d x  +
            \frac{E^2 + {E'}^2}{E E'}\, \frac{\Phi'(\xi)}{\xi} \right) 
\end{equation}
where
\begin{equation}
  \xi\equiv 
        \left(\frac{\omega}{\kappa E'}  \right)^{2/3}
    =    \left(\frac{u}{\kappa (1-u)}  \right)^{2/3} \quad .
  \label{eq:DiffEnerProbEmissSrxi}
\end{equation}
\end{subequations}
The result in Eqs.~(\ref{eq:DiffEnerProbEmissSr}) agrees with
previous calculations~\cite{Be:82,Sokolov:nk,Ba:88,Ka:96}.

Other important quantities in the study of the electromagnetic shower in
magnetic field are the differential energy loss for unit time and 
unit of photon energy:
\begin{equation}
 \label{eq:DiffFracEnerEmissSr}
 \frac{dE}{dtdu} = - \frac{\alpha}{\hbar} (mc^2)^2\, u 
                     \left( \int^{\infty}_{\xi} \Phi(x)d x  +
              \frac{1+ (1-u)^2}{1-u} \frac{\Phi'(\xi)}{\xi} \right) \quad ,
\end{equation}
where $\xi$ is given in Eq.~(\ref{eq:DiffEnerProbEmissSrxi}), and the
total energy loss per unit time (emissivity):
\begin{equation}
   \frac{dE}{dt} = - \frac{\alpha}{\hbar}\, (mc^2)^2
                \frac{\kappa^2}{2} \int_0^\infty 
        \frac{4 + 5 \kappa x^{3/2} + 4 \kappa^2 x^3}
             {(1+\kappa x^{3/2})^4}   x   \Phi'(x) dx   \quad .
                 \label{eq:TotalEmissSr}
\end{equation}
Note that, as expected from the lack of other dimensional scales in the
limit of $E \gg m$, the spectral emissivity in 
Eq.~(\ref{eq:DiffFracEnerEmissSr}) depends only from $\kappa$ and $u$
(scaling), and the emissivity in Eq.~(\ref{eq:TotalEmissSr})
depends only from the characteristic parameter $\kappa$,
apart from the overall factor $(mc^2)^2$.

Before studying the angular dependence of the emitted photons let us
discuss the total energy loss as a function of $\kappa$,
Eq.~(\ref{eq:TotalEmissSr}), and the energy dependence of the differential
energy loss, Eq.~(\ref{eq:DiffFracEnerEmissSr}).
For $\kappa \ll 1$ the energy loss goes to zero as $\kappa^2$, therefore
we limit our discussion to the more physically important case
$\kappa \gtrsim 1$.

The main feature of Eqs.~(\ref{eq:AngDiffProbEmissSru}) and 
(\ref{eq:DiffFracEnerEmissSr}) which determines both the energy and the
angular dependence is the presence of the Airy function $\Phi(x)$ that
goes to zero exponentially for large values of $x$. Only for the
purpose of this discussion
we use $x \lesssim 1$ as a simple threshold value (note that
$\Phi(1)/ \Phi(0) = 0.381$, while $\Phi(2)/\Phi(0) = 0.0984$), {\em i.e.},
we assume for the purpose of this discussion that most of the photons be
emitted for value of $x \lesssim 1$
(the discussion does not change if we use 2 or 3 intestead of 1), and use the 
ultrarelativistic limit $2(1-\cos\vartheta)\sim \vartheta^2\sim (m/E)^2 \ll 1$.

\begin{figure}[p]
\psfig{figure=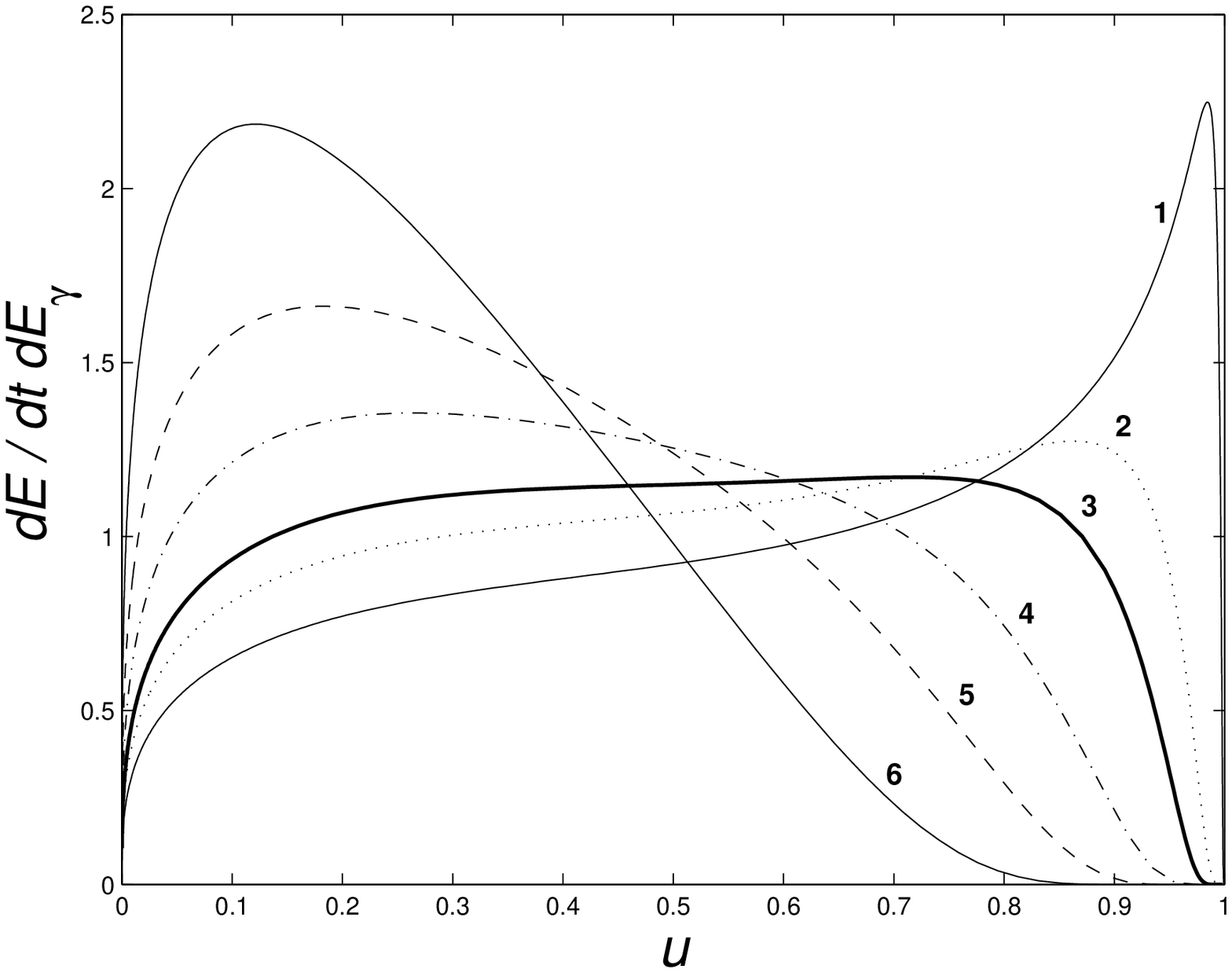,%
bbllx=40pt,bblly=190pt,bburx=540pt,bbury=590pt,%
height=15.0cm,angle=90}
%height=18cm}
%width=16.0cm,height=15.0cm,angle=90}
\caption{
Differential synchrotron-radiation emission probability as function of the 
fractional energy carried by the photon $u=\omega/E$,
Eqs.~(\ref{eq:DiffFracEnerEmissSr}), for several values of the 
characteristic parameter $\kappa=(H/H_c)(E/m)$: curve 1 (2, 3, 4, 5, 6)
corresponds to $\kappa= 100$ (10, 5, 2, 1, 0.5).
All curves are normalized such that the area is 1.
\label{Fig:SrFracEnerDist}
}
\end{figure}

According to this criterion the differential energy loss per unit
of photon energy,  Eq.~(\ref{eq:DiffFracEnerEmissSr}), is large when
$\xi \lesssim 1$, and, therefore, most of the photons are emitted with
a fractional energy $u$ that verify the condition
\begin{equation}
 u \lesssim \frac{\kappa}{1+\kappa} \quad ,
\end{equation}
{\em i.e.}, photons with a large fraction of the electron energy are emitted
only for relatively large values of $\kappa$. In addition, since the
energy loss is proportional to the energy fraction $u$ carried out by the
photon, the energy loss goes to zero with $u$.
Figure~\ref{Fig:SrFracEnerDist} summarizes this discussion by showing 
the probability of emission as function of the energy fraction with an
arbitrary normalization of 1 at $u=0$ for different values of $\kappa$.
Curve 1, which corresponds to the largest value of $\kappa=100$, has a peak
for values of $u$ close to 1, while curve 6, which corresponds to the
smallest value of $\kappa=0.5$, is peaked at values of $u$
close to 0.1 and goes to  zero at $u=0$. 

The same criterion applied to the angular distribution, {\em i.e.}
the constrain
\begin{equation}
\label{limitX}
X =  \left( 1+ \left(\vartheta\frac{E}{m}\right)^2 \right)
      \left(\frac{u}{(1-u)\kappa} \right)^{2/3} \lesssim 1
\end{equation}
implies that the angle within which most of the photons are emitted  
depends on the energy:
\begin{equation}
\label{limittheta}
\vartheta\frac{E}{m}  \lesssim 
      \sqrt{\left(\frac{(1-u)\kappa}{u} \right)^{2/3} - 1 }
      = \sqrt{1 / \xi -1 }
\end{equation}
when $\xi<1$. If instead $\xi \geq 1$ the angular distribution decays
exponential from the value $\vartheta E/m=0$ with a width proportional to
$\sqrt{\kappa(1-u)/u}$.
In other words photons with large energy fraction, $u \sim 1$, can be
emitted within a smaller angle than photons of low energy $u \ll 1$. 
Figure~(\ref{Fig:SrAngDist}) demonstrate this effect by showing 
the probability of emission as function of the variable  
$\sqrt{2(1-\cos\vartheta)} E/m \sim \vartheta E/m $ for four
choices of the couple $(u,\kappa)$. Dashed courves (1 and 3) have
$\kappa=0.1$ and show how the angle becomes more than an order of
magnitude larger going from $u=0.5$ (1) to $u=10^{-3}$ (3). The same
effect is show for  $\kappa=1$ going from curve 2 to curve 4.
The same angle grows more than a factor of two going from 
$\kappa=0.1$ 1 (3) to $\kappa=1$ 2 (4).

%%%%%%%%%%%%%%%%%%%%%%%%%%%%%%%%%%%%%%%%%%%%%%%%%%%%%%%%%%%%%%%%%%%%%%
\section{Pair production}
\label{sect3}
The amplitude for pair production can be obtained from the amplitude for
synchrotron radiation using the cross-channel symmetry~\cite{Be:82}.
Results mediated over the angular distribution can also be found
in~\cite{To:52,Ro:52,Erber:1966vv,Tsai:1974fa,Sokolov:nk,Ba:88,Ka:96}.

The calculation follows closely the steps in the previous Section with
the necessary formal differences. The characteristic parameter $\chi$ is
analogous to $\kappa$ in Eq.~(\ref{eq:CharacPar}) with the substitution
of the incoming-electron energy $E$ with the incoming-photon energy 
$\hbar \omega$:
\begin{equation}
     \chi =  \frac{H}{H_c} \frac{\hbar \omega}{mc^2} \quad :
     \label{eq:PpCharacPar}
\end{equation}
we use a different notation for clarity. Now the four-momenta of the
incoming photon and of the outcoming electron and positron are 
$K=(\omega , \mathbf k)$, $P_- = (E_- , \mathbf p_-)$, and
$P_+ = (E_+ , \mathbf p_+)$. As in the previous section,
we work in the limit when incoming and outgoing particles are
relativistic and the field is relatively low, $H\ll H_c$. 

After performing the cross-channel transformations
$(E,  \mathbf p ) \rightarrow (-E_+,  \mathbf -p_+)$, 
$(E',  \mathbf p') \rightarrow (E_-, \mathbf p_-)$, 
$(\omega , \mathbf k) \rightarrow (-\omega , -\mathbf k)$,
and substituting the emitted-photon phase space with the one of the
created positron (or electron), we obtain the analogous of
Eq.~(\ref{eq:EProbSrNp}) for the probability of pair production by an
unpolarized high-energy photon that propagates othogonal to the 
magnetic field  (the probability is summed over the final spins and
integrated over the electron direction, if we measure the positron energy:
the r\^{o}les of electron and positron can be exchanged).

This formula can again be expanded in the formation time $\tau_f$,
Eq.~(\ref{eq:FormationTime}), and the result is:
\begin{subequations}
\label{eq:PpSolidAngDiffProbEmissNatU}
\begin{equation}
\frac{dw}{dt dE_+ d\Omega_+}  = 
\frac{e^2}{\pi}
   \left[  \frac{({E_+}^2 + {E_-}^2){E_+}}{\omega^2 E_-} X
                +  \left( \frac{E_+^2}{\chi \omega E_-} \right)^{2/3}
   \right] \Phi(X) 
\end{equation}
where
\begin{equation}
X \equiv 2   \left( \frac{E_+}{m} \right)^2
\left( 1  - \left( 1-\frac{m^2}{2E_+^2} \right) \cos\vartheta \right) 
\left( \frac{\omega^2}{\chi(E_+ E_-)} \right)^{2/3} \quad .
\end{equation}
\end{subequations}

Again there is no dependence on the azimuthal
angle $\varphi$ and $d\Omega_{+} = 2\pi d\cos\vartheta$.
Introducing the fractional energy carried out by the positron,
$v = E_+ / \omega $, Equations~(\ref{eq:PpSolidAngDiffProbEmissNatU}) become
\begin{subequations}
\label{eq:AngDiffProbEmissPpGenCaseFracEn}
\begin{equation}
    \frac{dw}{dt dE_+ d\cos\vartheta}  =  
         \frac{2 \alpha}{\hbar} v
    \left[ \frac{v^2 + (1-v)^2}{(1-v)} X + v  
            \left( \frac{1}{v (1-v)\chi} \right)^{2/3}
    \right] \Phi(X)  
\end{equation}
with
\begin{eqnarray}
   X &=& 2 \left( \frac{E_+}{m} \right)^2
        \left( 1  - \left( 1-\frac{m^2}{2E_+^2} \right) \cos\vartheta \right) 
        \left(\frac{1}{v (1-v)\chi} \right)^{2/3}  \\
     &=&    \left( 1+2(1-\cos\vartheta) \frac{E_+^2}{m^2} -(1-\cos\vartheta)
           \right)
      \left(\frac{1}{ v (1-v)\chi} \right)^{2/3} \quad ,
\end{eqnarray}
\end{subequations}
and $2(1-\cos\vartheta)\sim \vartheta^2 \sim (m/E)^2$ the convenient
angular variable in the ultrarelativistic limit.

As in the case of the synchrotron radiation, we can cross-check our result
by integrating Eq.~(\ref{eq:AngDiffProbEmissPpGenCaseFracEn}) over the polar
angle and comparing the differential emission probability per unit
energy with Refs.~\cite{Erber:1966vv,Be:82,Tsai:1974fa,Sokolov:nk,Ba:88,Ka:96}:
\begin{subequations}
\label{eq:PpDiffEnerProbEmissNatU}
\begin{equation}
\frac{dw}{dt dE_+} = 
 e^2 \left(\frac{m}{\omega}\right)^2 
   \left( \int^{\infty}_{\xi} \Phi(x)d x - 
           \frac{E_+^2 + E_-^2}{E_+ E_-} \frac{\Phi'(\xi)}{\xi}  
  \right)
\end{equation}
where
\begin{equation}
\label{eq:PpDiffEnerProbEmissNatUx}
\xi  \equiv \left(\frac{1}{\chi} \frac{\omega^2}{E_+ E_-} \right)^{2/3} 
 = \left(\frac{1}{\chi v (1-v)} \right)^{2/3}   \quad .
\end{equation}
\end{subequations}

The differential emission probability per unit time and unit 
of fractional energy $v$ is:
\begin{equation}
\label{eq:PpDiffFracEnerProbEmiss}
 \frac{dw}{dt dv} = 
             \frac{\alpha}{\hbar}   \frac{(mc^2)^2}{\hbar\omega} 
       \left( \int^{\infty}_{\xi} \Phi(x)d x  - 
                  \frac{v^2 + (1-v)^2}{v(1-v)}\, \frac{\Phi'(\xi)}{\xi}  
       \right)
\end{equation}
with  $\xi$ defined in Eq.~(\ref{eq:PpDiffEnerProbEmissNatUx}), while the
total pair production probability is:
\begin{equation}
 \label{eq:PpTotalProb}
\frac{dw}{dt} =  - \frac{\alpha}{\hbar}  \frac{(mc^2)^2}{\hbar\omega} 
      \int_{ \left(4 / \chi \right)^{2/3}}^\infty
  \left( \frac{ 2 x^{3/2}  + 1/ \chi}{x^{11/4}(x^{3/2} - 4/ \chi)^{1/2}} 
 \right) \Phi'(x)  d x  \quad .
\end{equation}

Note again that the differential emission probability depends only 
from $\chi$ and $v$ and the  total pair production probability only from
$\chi$ (scaling), apart the dimensional constant in front $(mc^2)^2$, as
a consequence of the lack of other dimensional scales in the limit of
$E \gg m$. Note also that the pair production probability, differently
from the synchrotron emission, is  exponentially suppressed in the limit
of $\chi \ll 4$, since the Airy function decays exponentially for large
values of its argument. The physical cause of this suppression is
the presence of a threshold for pair creation. In the following
discussion we consider the range  $\chi \gtrsim 4 $.

As in the case of the synchrotron radiation the main feature of 
Eqs.~(\ref{eq:AngDiffProbEmissPpGenCaseFracEn}) and 
(\ref{eq:PpDiffFracEnerProbEmiss}) that determines both the energy and the
angular dependence is the exponential suppression of the Airy function 
$\Phi(x)$ with growing $x$; we use the threshold value $x \lesssim 1$
and work in the ultrarelativistic limit $\vartheta \sim m/E \ll 1$.
In addition the pair creation probability is symmetric in the two variable
$E_+$ and $E_-$, {\em i.e.}, is symmetric respect to the point $v=1/2$.

\begin{figure}[p]
\psfig{figure=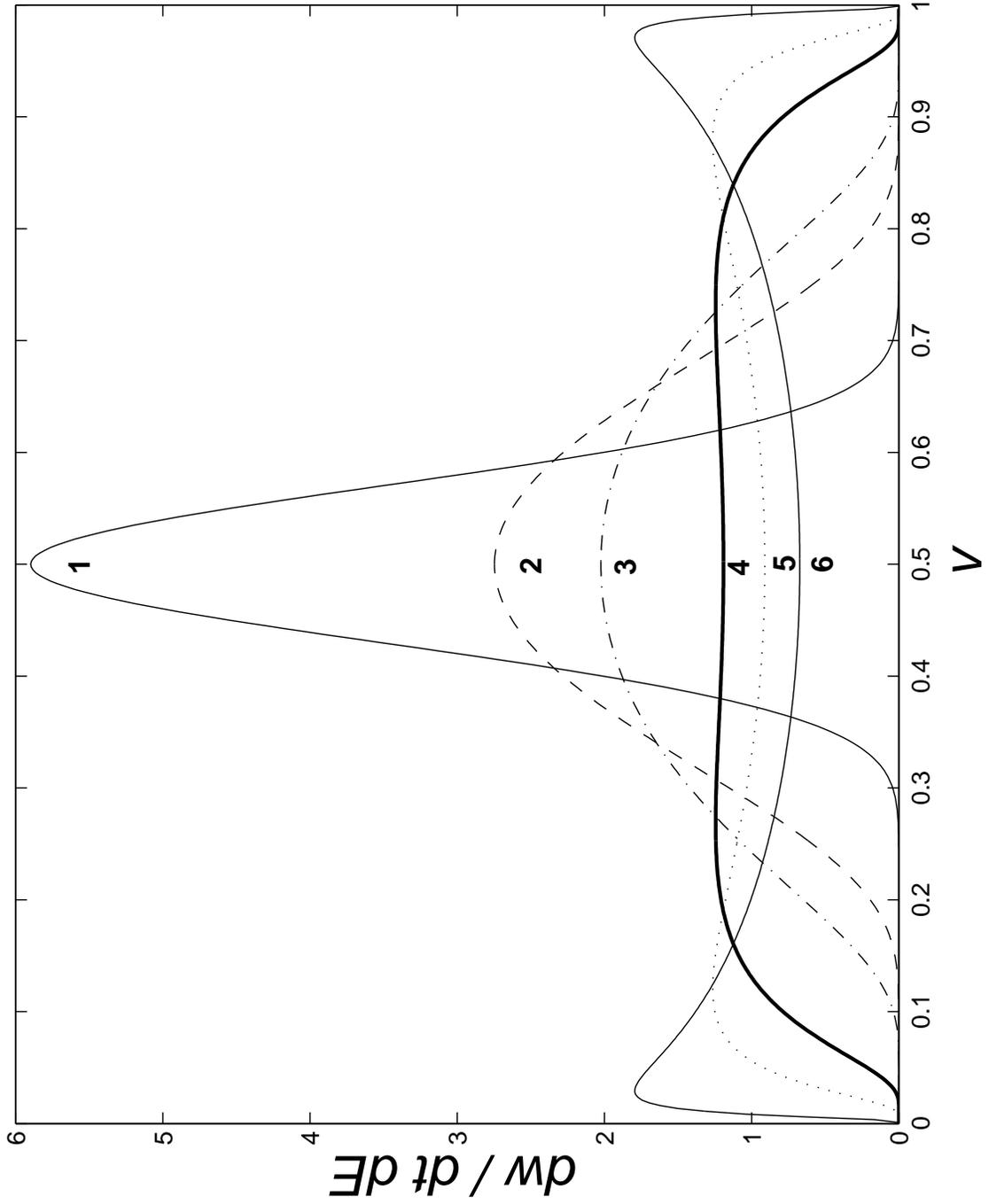,%
bbllx=60pt,bblly=190pt,bburx=545pt,bbury=590pt,%
height=15.0cm,angle=90}
\caption{
Differential pair-production emission probability  as function of
the fractional energy carried by the electron/positron
$v$, Eqs.~(\ref{eq:PpDiffFracEnerProbEmiss}), for several values of
the characteristic parameter $\chi=(H/H_c)(\omega/m)$: curve 1 (2, 3, 4, 5, 6)
corresponds to $\chi= 0.1$ (0.5, 1, 4, 10, 100).
All curves are normalized such that the area is 1.
 \label{Fig:PpFracEnerDist}
}
\end{figure}

In this case the chosen criterion imples that the differential energy
loss per unit of electron/positron energy,  
Eq.~(\ref{eq:PpDiffFracEnerProbEmiss}), is large when 
$\xi \lesssim 1$, and, therefore, most of the $e^+ e^-$ are emitted with
a fractional energy $v$ that verify the condition
\begin{equation}
 -\sqrt{\frac{1}{4}-\frac{1}{\chi}} 
         \lesssim v-\frac{1}{2} \lesssim 
  \sqrt{\frac{1}{4}-\frac{1}{\chi}} \quad ,
\end{equation}
when $\chi>4$, {\em i.e.}, $e^+ e^-$ can be emitted with one of them
carrying a large fraction of the photon energy only for relatively large
values of $\chi$. 
Figure~\ref{Fig:PpFracEnerDist} summarizes this discussion by showing 
the probability of emission as function of the energy fraction with the area
arbitrary normalized to 1 for different values of $\chi$. The smallest value
of $\chi=0.1$ (curve 1) is strongly peaked at $v=1/2$, while the largest
value of $\chi=100$ (curve 6) has a much flatter distribution with peaks
at values close to $v=0$ and $v=1$, in spite of the fact that the
distribution must go to zero at exactly $v=0$ and $v=1$.

The angular distribution is described by exactly the same constrain
found for the synchrotron radiation
\begin{equation}
\label{limitXee}
X =  \left( 1+ \left(\vartheta\frac{E}{m}\right)^2 \right)
      \left(\frac{1}{v(1-v)\chi} \right)^{2/3} \lesssim 1 \quad ,
\end{equation}
which implies that the energy-dependent angle within which most of the
pairs are emitted is:  
\begin{equation}
\label{limitthetaee}
\vartheta\frac{E}{m}  \lesssim 
      \sqrt{\left(v(1-v)\chi \right)^{2/3} - 1 }
      = \sqrt{1 / \xi -1 }
\end{equation}
where we are considering only values of $\xi<1$. 
If instead $\xi \geq 1$ the angular distribution decays
exponential from the value $\vartheta E/m=0$ with a width proportional to
$\sqrt{(1-v)v\chi}$. In other words
pairs  that share equally their energy $v \sim 1/2$
can be emitted within a larger angle than pairs where one of the two
particles carries a large part of the energy, $v \sim 1$ or $v \sim 0$.
Figure~(\ref{Fig:PpAngDist}) demonstrates this effect by showing 
the probability of emission as function of the variable  
$2(1-\cos\vartheta) (E/m)^2 \sim \vartheta^2 $; the distribution has been
arbitrary normalized such that it is equal to 1 at $\vartheta E/m = 10^{-1}$.
Going from an asymmetric distribution of the energy, $v=0.1$, (curves 1 and 3)
to a symmetric one, $v=0.5$, (curves 2 and 4) the angle becomes wider; the
same happens when $\chi$ increases from 4 (curves 1 and 2) to 100
(curves 3 and 4).

%%%%%%%%%%%%%%%%%%%%%%%%%%%%%%%%%%%%%%%%%%%%%%%%%%%%%%%%%%%%%%%%%%%%%%%%%
\section{Conclusions}
\label{sect4}

In this paper we have studied the angular dependence of photons emitted
by UHE electrons and of the $e^+ e^-$ pairs emitted by UHE photons in
a static magnetic field: this dependence is needed for detailed studies
of the electromagnetic cascade in magnetic fields, such as those initiated
by UHE cosmic rays in the geomagnetic field, or by charged particles emitted
by pulsar.

The main results are shown in Eqs.~(\ref{eq:AngDiffProbEmissSru}) for the
magnetic
bremsstrahlung and in Eqs.~(\ref{eq:AngDiffProbEmissPpGenCaseFracEn}) for
the  pair production. For simplicity we have sketched the derivation of
these formulae in the case of propagation in the plane orthogonal to the
magnetic field, but it is possible to show, as we have explicitely verified
by performing the appropriate
Lorentz transformations, that the same formulae are valid in the general
case if $H$ is substituted with $H_{\perp}$, the component of the magnetic
field perpendicular to the propagation.

\begin{figure}[p]
\psfig{figure=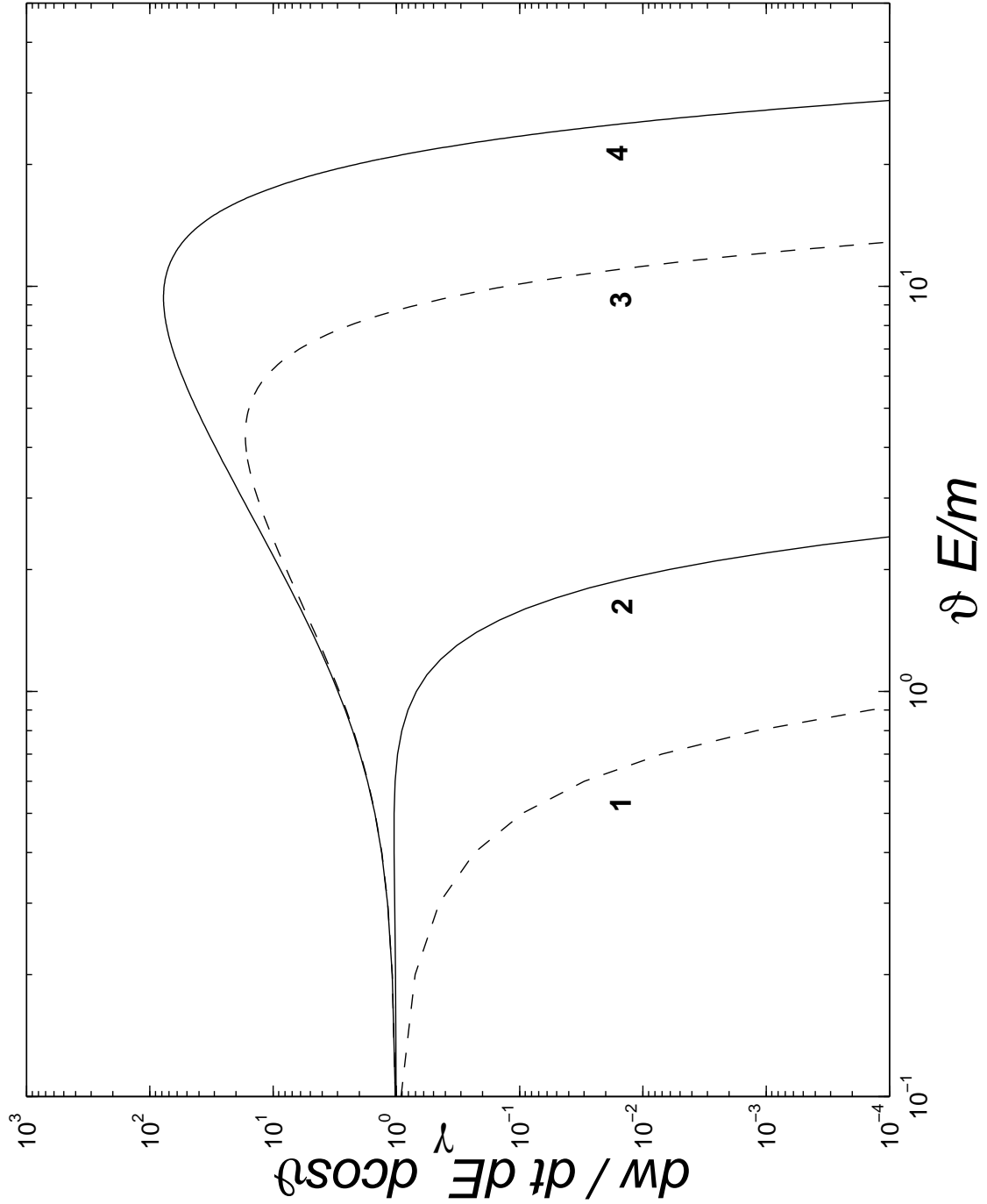,%
bbllx=40pt,bblly=185pt,bburx=540pt,bbury=590pt,%
height=15.0cm,angle=90}
\caption{
Angular distribution of the synchrotron-radiation emission probability,
Eqs.~(\ref{eq:AngDiffProbEmissSru}), for two values of the
fractional energy carried by the photon $u=\omega/E$ and of the
characteristic parameter $\kappa$: curve 1 corresponds to
$u=0.5$ and $\kappa=0.1$;  curve 2 to $u=0.5$ and $\kappa=1$; 
curve 3 to $u=10^{-3}$ and $\kappa=0.1$;
curve 4 to $u=10^{-3}$ and $\kappa=1$. 
The angular variable is $\sqrt{2(1-\cos\vartheta)}E/m \sim \vartheta E /m$;
the normalization is such that curve is 1 at 
$\vartheta = 0$.
 \label{Fig:SrAngDist}
}
\end{figure}

\begin{figure}[p]
\psfig{figure=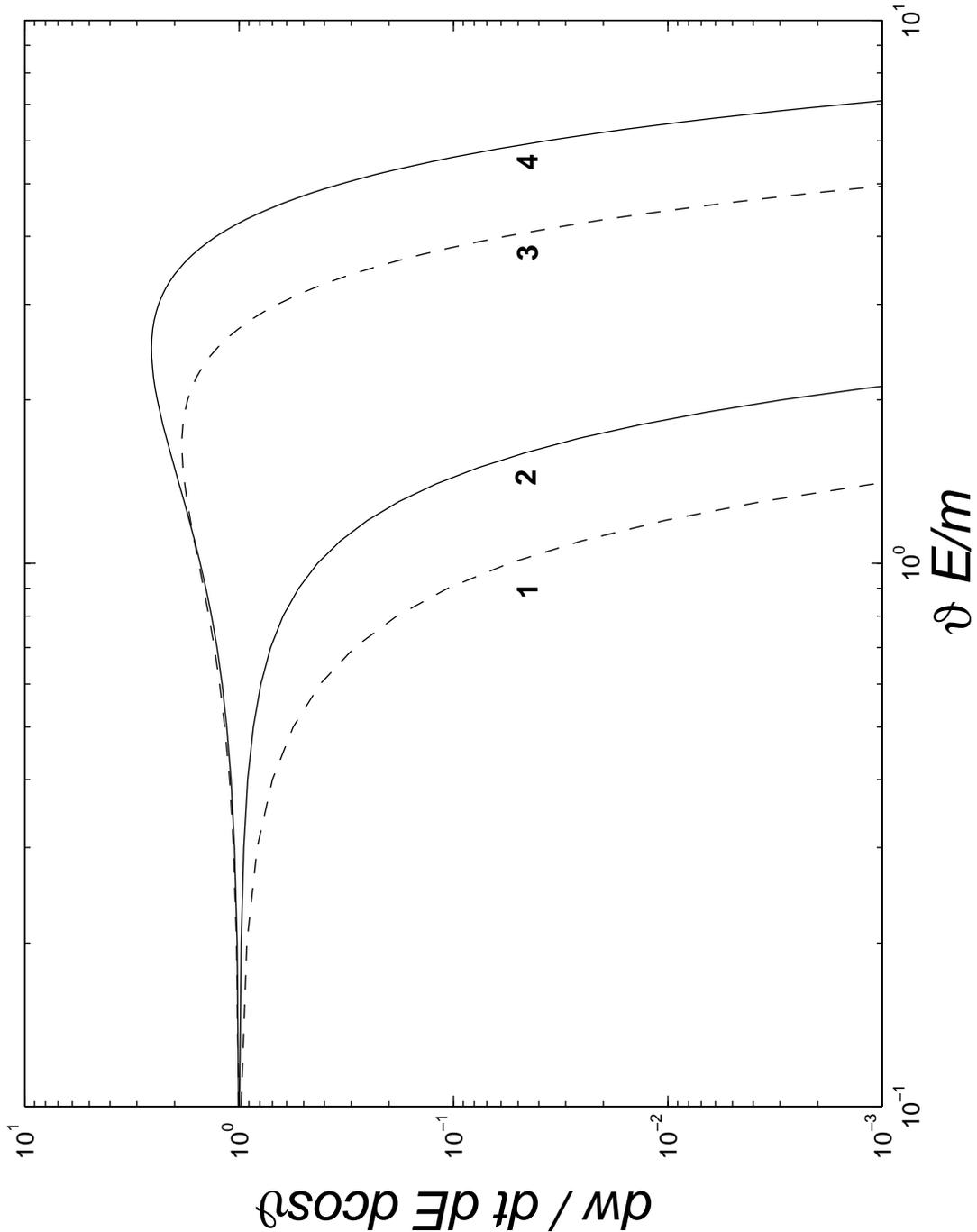,%
bbllx=35pt,bblly=185pt,bburx=550pt,bbury=590pt,%
height=14.5cm,angle=90}  
\caption{
Angular distribution of probability of emitting a positron (electron) 
with fractional energy $v$ by a UHE photon,
Eqs.~(\ref{eq:AngDiffProbEmissPpGenCaseFracEn}),
for two values of $v$ and of the characteristic parameter $\chi=1$:
curve 1 corresponds to
$v=0.1$ and $\chi=4$;  curve 2 to $v=0.5$ and $\chi=4$; 
curve 3 to $v=0.1$ and $\chi=100$;
curve 4 to $v=0.5$ and $\chi=100$.
The angular variable is $\sqrt{2(1-\cos\vartheta)}E_{+}/m 
\sim \vartheta E_{+} /m$; the normalization is such that curve is 1 at 
$\vartheta = 0$.
  \label{Fig:PpAngDist}
}
\end{figure}

These results are also plotted as function of the angle for different
values of the fractional energy in Figs.~\ref{Fig:SrAngDist} and 
\ref{Fig:PpAngDist} and briefly discussed in text. The angle becomes
wider for larger values of the characteristic parameter and for
smaller energy fraction (synchrotron radiation) or more symmetric energy
fraction (pair production).

We have verified that our results, when integrated over the emission angle,
reproduce the known results for the differential in the energy and total 
probability of emission.

%%%%%%%%%%%%%%%%%%%%%%%%%%%%%%%%%%%%%%%%%%%%%%%%%%%%%%
\begin{acknowledgments}
This work is partially supported by M.I.U.R. (Ministero dell'Istruzione,
dell'Universit\`a e della Ricerca): ``Cofinanziamento'' P.R.I.N. 2001.
\end{acknowledgments}
%%%%%%%%%%%%%%%%%%%%%%%%%%%%%%%%%%%%%%%%%%%%%%%%%%%%%%%%%%%%%%%%%%%%%

%

\end{document}